# Extinction of megafauna: How could the research get so wrong?

Ron W. Nielsen[1]

**Abstract**. Published evidence for the human-mediated extinction of megafauna is examined and is found to be unsubstantiated. It is shown that the claimed evidence is not based on data describing the growth of human population but on the fabricated data. However, even these fabricated data, which were claimed to support the human-induced extinction of megafauna, contradict this claim. The belief in the human-induced extinction of megafauna appears to be so strong that even contradicting evidence based on the fabricated data is interpreted as the evidence supporting this belief.

## Introduction

In terrestrial zoology, megafauna is made of large animals (Farina, Vizcaino & De Iuliis, (2013); Hube, Sinclair & Lumpe, 2005; Martin & Klein, 1984; MacPhee & Sues, 1999; Stuart, 1991). In soil science, megafauna is made of small animals such as moles, mice, hares, rabbits, gophers, snakes, and lizards (Raffery, 2011). I am discussing here megafauna as defined in terrestrial zoology. In particular, I am discussing just one claimed evidence for the human-induced extinction of megafauna. This is a disturbing example because it shows how even the most obviously incorrect claim can be published in a peer-reviewed journal. It also shows how the obviously fabricated "data" were published in another peer-reviewed journal. How is it possible to publish unscientific papers in scientific journals? How much trust can we have in the peer-reviewing system?

I am also showing that even these fabricated data do not support the claim of human-mediated extinction of megafauna. Arguments in favour of this claim are so obviously incorrect that it is surprising that they were ever published. How could science get it all so wrong?

This discussion was first published in 2013. The new version (2017), presented here, contains new information about the growth of population.

## Claimed empirical evidence

An excellent way of showing that extinction of megafauna might have been caused by humans would be to show a clear and convincing correlation between the extinction of megafauna and the growth of human population. We would like to see a clear boosting in the growth of human population correlated with the decline in the population of megafauna. Such a correlation would not necessary prove that the boosting in the growth of population was *caused* by the extinction of megafauna but it would be an interesting evidence, which could contribute to solving the puzzle of this extinction.

Such a study, or at least the attempt of such a study, was published by Barnosky (2008). His results are reproduced in Figure 1. They have been claimed to show a close correlation between the rapid decline in the number of species of megafauna and the growth of human population. "The numbers of megafauna species lost were modest until the human growth curve *began its rapid exponential rise* between 15.5 and 11.5 kyr B.P." (Barnosky, 2008, p. 11544; emphasis added.). Thus, the extinction pulse (the sudden and relatively short-lasting decline in the number of species) is explained as being correlated with the beginning of the rapid increase in the size of human population.

---

[1] AKA Jan Nurzynski, Environmental Futures Research Institute, Gold Coast, Griffith University, Qld, 4222, Australia, ronwnielsen@gmail.com

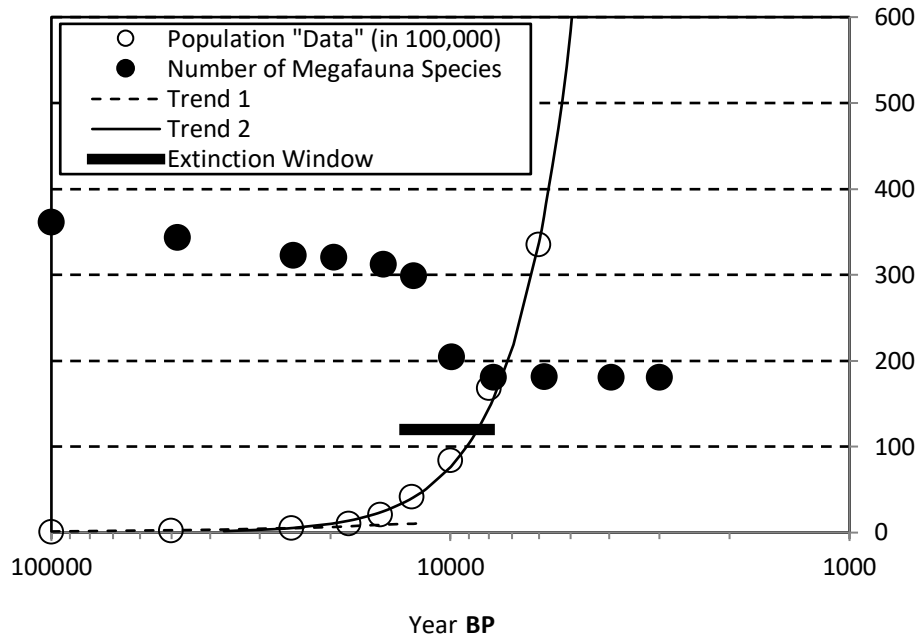

**Figure 1.** *The claimed but non-existent correlation between the extinction of megafauna and the growth of population. (Barnosky, 2008, p. 11544). There is no correlation because the discontinuity in the number of species of megafauna is not reflected in any discontinuity in the growth of population.*

And here we have the first and glaring problem, which should have been noticed by Barnosky and by the referee of this scientific journal: there is obviously *no correlation* between the growth of population and the extinction of megafauna. The distribution describing the extinction of megafauna is clearly *non-monotonic*. It is characterised by a clear and sudden decline, a clear and sudden change of direction, a clear sudden discontinuity. However, the growth of population is *monotonic*. It is not characterised by a clear sudden boosting, a clear sudden change of direction or a clear sudden discontinuity at the time of the intensified extinction.

Furthermore, the question is why the extinction spree lasted for only such a short time. Why was the process of extinction discontinued even though the claimed size of population continued to increase, the hunting ability was no doubt improving and the number of species was still high, reduced from around 300 just before the claimed extinction pulse to around 180 towards its end? Why did humans lost interest in hunting megafauna and in using this productive source of food to support their growth? Maybe the extinction of megafauna had nothing to do with human activities.

It is interesting how much one can learn sometimes by simply looking at the data from a different perspective. We often do not need to carry out some complicated analysis of data. Just display the data in a different way and you can discover features, which were not so clear in another display. Figure 2 presents precisely the same data as published by Barnosky (2008) and as shown in Figure 1 but now displayed using the logarithmic scales of reference for both axes. In this representation, population data follow approximately two straight lines, both being clearly *uncorrelated* with the extinction data. *The claimed correlation does not exist.*

The "data" presented in Figure 2 show that there was a beginning of a rapid increase in the growth of population around 25 ka (thousands of years ago). However, there was no change in the growth of population at the beginning of the extinction pulse at around 15.5 ka. The two distributions, one describing the growth of population and the other describing the extinction of megafauna, are clearly *not correlated*. A 10,000-year gap between the beginning of the rapid increase in the growth of population around 25 ka and the beginning of the extinction pulse around 15.5 ka is too large to claim any meaningful correlation between these two distributions.

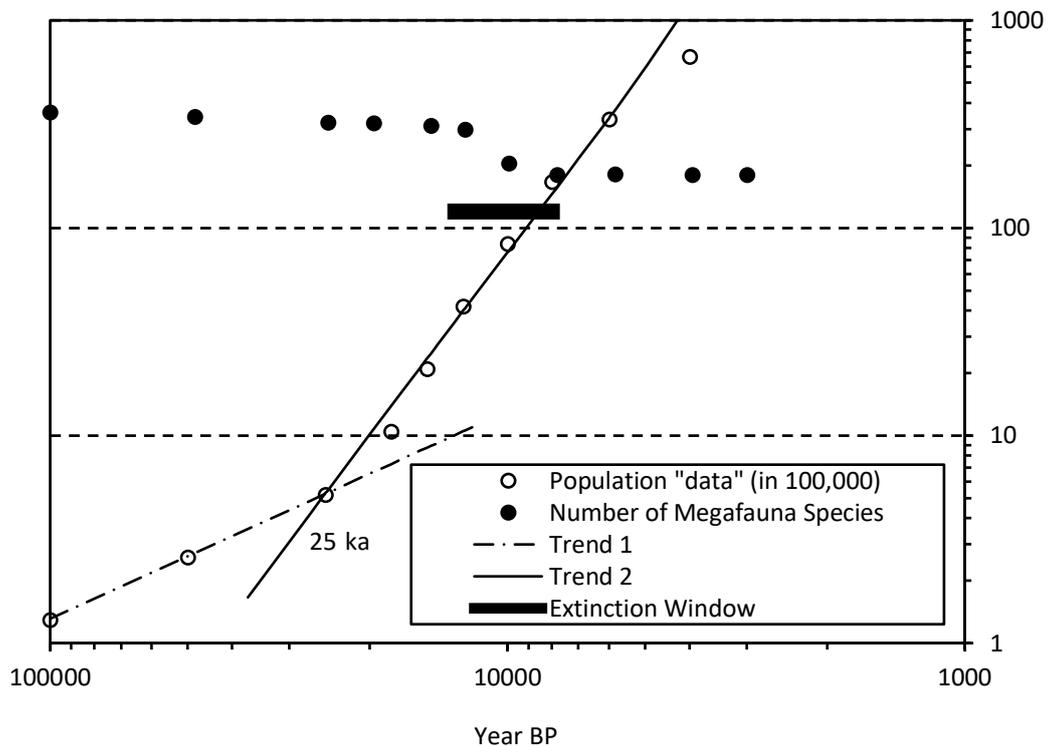

**Figure 2.** *The evidence claimed to be in support of the anthropologically driven extinction of megafauna (Barnosky 2008), shown earlier in Figure 1, is now displayed using logarithmic scales of reference.* There is no correlation *between the growth of population and the extinction of megafauna. The growth of population was monotonic at the time of the extinction of megafauna. The growth of population was not boosted by this extinction. There is a clear decline in the number of species of megafauna but this decline is not matched by a surge in the growth of population.*

Results published by Barnosky (2008) are interesting but for the unintended reason. *He has demonstrated that the extinction of megafauna is not correlated with the growth of population* and consequently, he has shown that the extinction of megafauna was not caused by humans, or at least he has demonstrated that *there is no supporting evidence for claiming that the extinction of megafauna was caused by humans*. The hypothesis of the human-assisted extinction is not supported by the population "data."

However, the published claim is even more intriguing because the population "data" used to support the hypothesis of the human-mediated extinction of megafauna *do not even describe the growth of human population*. These "data" were created by Hern (1999) and are reproduced in Table 1. Even a quick glance at this table shows that the published set of numbers cannot be interpreted as describing the growth of human population. How was it possible that Barnosky did not see it? How was it possible that these obvious signs of deliberate fabrication of data were not noticed by the referees of both scientific journals, the journal that published Hern's "data" and the journal that published Barnosky's results?

*First*, genus *Homo* did not exist 3 Ma (million years ago). The set of numbers created by Hern (1999) cannot be linked with *Australopithecus* or with genus *Homo* because *Australopithecus* ceased to exist approximately 1.2 Ma and genus *Homo* did not emerge until around 2.5 Ma or even later if the questionable classification of *H. habilis* is excluded (see Figure 3). This set of numbers describes, therefore, the growth of a non-existing genus, which for convenience of this discussion we might call

*Phasmapithecus*, from Latin *phasma* or Ancient Greek *φάσμα* (ghost, spectre, apparition) and the Greek word *πίθηκος* (ape).

**Table 1.** *Population "data" created by Hern (1999) and used by Barnosky (2008) in support of the hypothesis of the anthropologically driven extinction of megafauna.*

| Year [BP] | Population | $T_2$ | Year [BP] | Population | $T_2$ |
|---|---|---|---|---|---|
| 3,000,000 | $2^0=1$ | 500,000 | 10,000 | $2^{23}=8,388,608$ | 2,000 |
| 2,500,000 | $2^1=2$ | 500,000 | 8,000 | $2^{24}=16,777,216$ | 2,000 |
| 2,000,000 | $2^2=4$ | 250,000 | 6,000 | $2^{25}=33,554,432$ | 2,000 |
| 1,750,000 | $2^3=8$ | 250,000 | 4,000 | $2^{26}=67,108,864$ | 1,000 |
| 1,500,000 | $2^4=16$ | 125,000 | 3,000 | $2^{27}=134,217,720$ | 700 |
| 1,250,000 | $2^5=32$ | 125,000 | 2,000 | $2^{28} \approx 268,434,000$ | 1650 |
| 1,000,000 | $2^6=64$ | 100,000 | 340 | $2^{29} \approx 536,868,000$ | 200 |
| 900,000 | $2^7=128$ | 100,000 | 140 | $2^{30} \approx 1,073,736,000$ | 80 |
| 800,000 | $2^8=256$ | 100,000 | 68 | $2^{31} \approx 2,147,472,000$ | 46 |
| 700,000 | $2^9=512$ | 100,000 | 22 | $2^{32} \approx 4,294,944,000$ | 37 |
| 600,000 | $2^{10}=1,024$ | 100,000 | 0=AD 1998 | $2^{32.5} \approx 6,000,000,000$ | |
| 500,000 | $2^{11}=2,048$ | 100,000 | | | |
| 400,000 | $2^{12}=4,096$ | 100,000 | | **Projections** | |
| 300,000 | $2^{13}=8,192$ | 50,000 | **AD 2053** | **Population** | $T_2$ |
| 250,000 | $2^{14}=16,384$ | 50,000 | 2013 | 8,589,888,000 | 40 |
| 200,000 | $2^{15}=32,768$ | 50,000 | 2053 | 16,777,216,000 | 50 |
| 150,000 | $2^{16}=65,536$ | 50,000 | 2103 | 34,359,552,000 | 70 |
| 100,000 | $2^{17}=131,072$ | 50,000 | 2173 | 68,719,104,000 | 70 |
| 50,000 | $2^{18}=262,144$ | 25,000 | 2243 | 137,438,208,000 | 70 |
| 25,000 | $2^{19}=524,228$ | 5,000 | 2313 | 274,876,416,000, | 70 |
| 20,000 | $2^{20}=1,048,576$ | 5,000 | 2383 | 549,752,832,000 | 70 |
| 15,000 | $2^{21}=2,097,152$ | 2,500 | 2453 | 1,000,000,000,000 | 70 |
| 12,500 | $2^{22}=4,194,304$ | 2,500 | | | |

$T_2$ – doubling time

*Second*, the created genus is endowed with bizarre reproductive properties and with the exceptional longevity, because it had an ability of self-replication after 500,000 years. (The first and only representative of this mysterious genus self-replicated after 500,000 years to increase its population from 1 in 3,000,000 BP to 2 in 2,500,000 BP.) Hern appears to be aware of this problem but he promptly points out that he could have started his construction from "Adam and Eve" (Hern, 1999, p. 63) confirming that the numbers he listed were obviously of his own creation and that they had nothing to do with the paleontological or archaeological research. Barnosky should have seen this comment. However, even if we start with two representatives of this phantom genus we still have the problem of longevity and of the peculiar procreation abilities because the first two representatives increased to four after 500,000 years. They were either endowed with the exceptional longevity or they were on the verge of extinction for at least 500,000 years. In fact, they were on the verge of extinction for at least 2,000,000 years because their population increased from 1 in 3,000,000 BP to only 64 in 1,000,000 BP.

*Third,* the set of numbers listed in Table 1 represents a crudely constructed patchwork of exponential growth, indicating again that it does not represent data but that it was fabricated. This set of numbers is characterised by drastically different doubling times and consequently by drastically different growth rates, showing no signs of smooth transitions between adjacent sections. Thus, for instance, around 2 Ma, the doubling time changed abruptly from 500,000 years to 250,000 years, decreasing neatly by a factor of two. This pattern of abrupt changes, often involving the factor of two or producing round numbers for the doubling times, continues until 3 ka when the doubling times stopped following the earlier regular pattern probably to force the created numbers to agree more closely with the data for the growth of human population during the AD era. However, for no apparent reason, from 2103 until 2453 the population of this phantom genus follows an exponential trajectory corresponding to the growth rate of 1%.

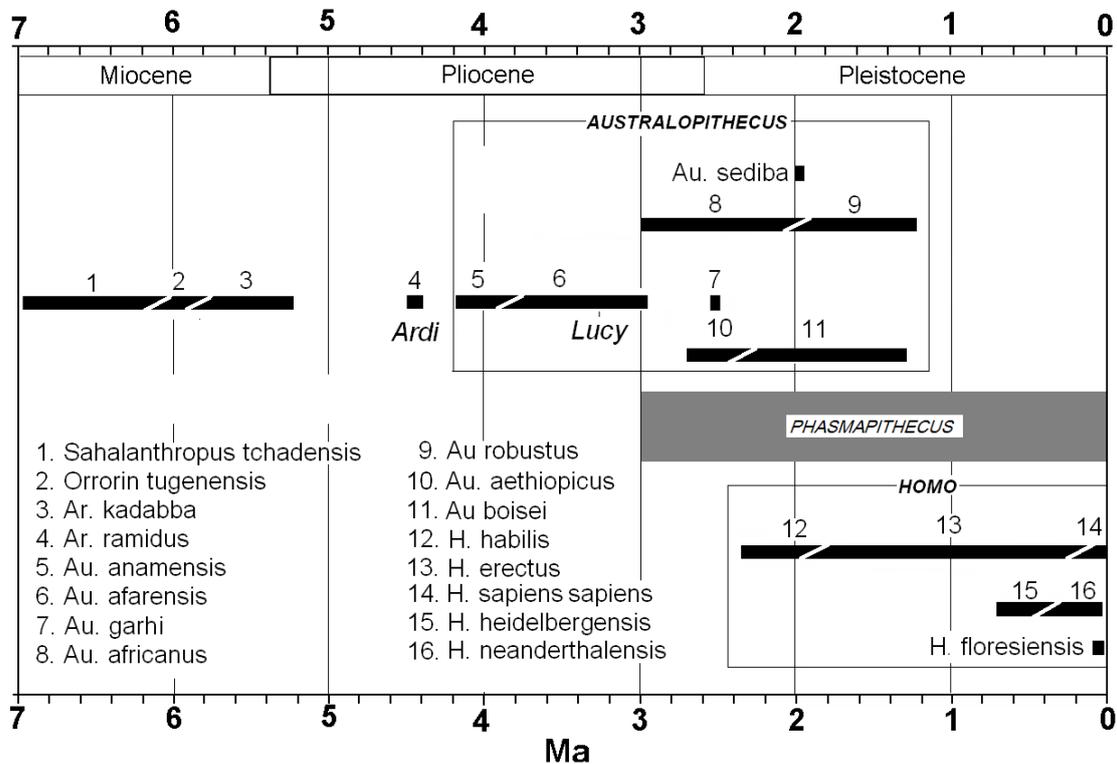

**Figure 3.** *The sequential representation of human evolution based on the data presented by Cremin (2007), Gibbons (2009), Heile-Selassie (2010), Thanukos (2010) and Zagura (2008) showing also the phantom genus,* Phasmapithecus, *created by Hern (1999) and used by Barnosky (2008) in support of the postulated human-assisted extinction of megafauna.*

*Fourth*, the projected population of *Phasmapithecus* (which is supposed to represent humans) is 1,000 billion in 2453, clearly showing that the prediction is unacceptably high. The latest predictions for the growth of population are 9.8 billion in 2050, around 12 billion in 2100 and the asymptotic maximum of 15.6 billion in a more distant future (Nielsen, 2017a).

All these features should have served as warning signs that the numbers published by Hern (1999) do not describe the growth of human population. How is it possible that Barnosky did not see these obvious signs of the deliberate fabrication of data? How is it possible that he did not notice Hern's comment that he could have started the construction of his "data" from Adam and Eve (Hern, 1999, p. 63)?

I have contacted Hern asking him for the explanation of his numbers and this is what he wrote. "I did not intend in any way to offer my table of the number of doublings as 'data' or a prediction. All of the numbers - and I don't think there are any exceptions - are imprecise to wildly wrong and based on estimates ranging from considered evaluation of available data (?) to total speculation" (Hern, 2013).

Hern explained that his sole purpose was to use these numbers to argue that in analogy with the growth of cancer cells, the size of human population might have already doubled a critical number of times (Hern, 2013). However, the calculation of the number of doubling times depends only on just two parameters, the initial and the final size of population or more generally on the initial and final size of any growing entity. It does not matter whether the growth is exponential or not, or made of exponential sections as used by Hern (1999) who has a strong preference for using exponential functions (Hern, 2013). He did not have to go through all this trouble of constructing a complicated table of numbers, which were supposed to represent the growth of human population. All he had to do was to assume the initial and the final sizes of population. Why to make it all so complicated if it is such a simple mathematical exercise?

The important point is that his *constructed "data" should have never been used in support of the hypothesis of the human-assisted extinction of megafauna* because they were fabricated and because they do not describe the growth of human population. However, as already pointed out earlier, *even these "data" do not show any correlation with the extinction pulse*.

It might be interesting to notice that the growth of *Phasmapithecus* was *boosted seven times* before the beginning of the massive extinction of megafauna in 15.5 ka but all these changes were ignored by Barnosky. It is also interesting that the growth of *Phasmapithecus* was *not* boosted at the time of the extinction pulse but now it was claimed in order to support the concept of the human-driven extinction. It appears that the desire to see a support for a cherished hypothesis can be so strong that it interferes with the scientific process of investigation.

Another "evidence" used by Barnosky (2008) in support of the human-mediated extinction of megafauna is the claimed close trade-off between the biomass of megafauna and the biomass of humans. Barnosky's calculations (Barnosky, 2008, p. 11455) are reproduced in Figure 4. "The results suggest that *biomass loss* from the early megafauna extinctions in Australia and the first pulse of extinctions in Eurasia and Beringia *were almost exactly balanced by the gain in human biomass*" (Barnosky, 2008, p. 11544; emphasis added.). This claim was endorsed and reinforced by Avise, Hubbell and Ayala (2008, p. 168) who accept that there is an "inverse relationship between human biomass and nonhuman megafaunal biomass".

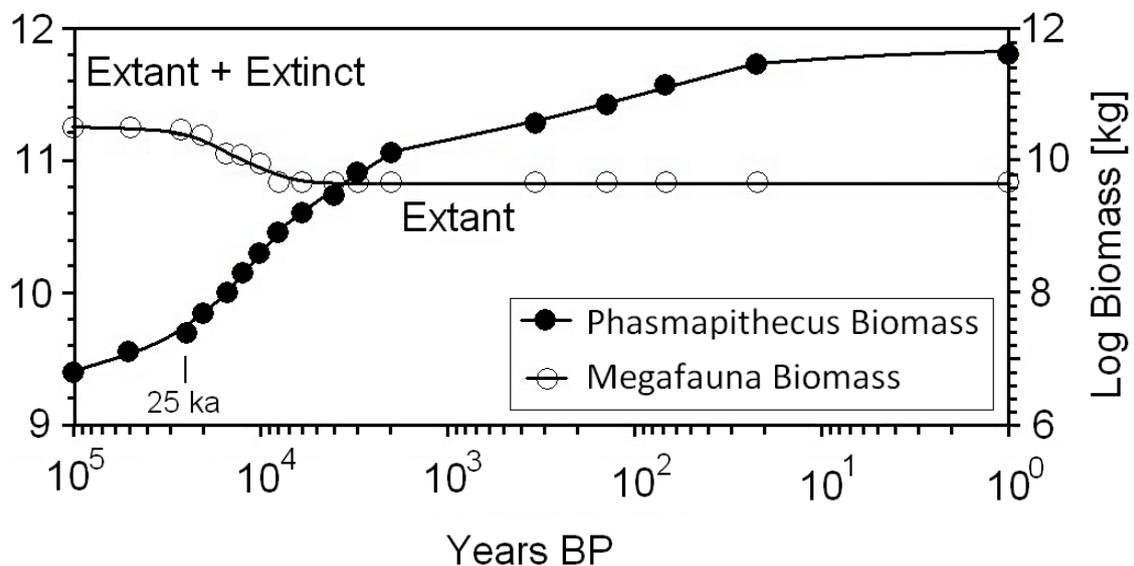

**Figure 4.** *The biomass trade-off between megafauna and the alleged human population, calculated by Barnosky (2008) and used in support of the hypothesis of the human-assisted extinction of megafauna. The alleged human population was not human population but the population of a phantom genus,* Phasmapithecus, *which never existed but which is described by the "data" fabricated by Hern (1999).*

The irony of it all is that it is not the human biomass but an imaginary biomass of an imaginary genus *Phasmapithecus*. What Barnosky represents as human biomass in his original figure is in fact the phantom biomass of a phantom genus. It is a non-existing biomass. The biomass of megafauna was simply disappearing. Consequently, Barnosky's claim of the close trade-off makes no sense. However, even it were human biomass it would be still hard to explain this close correlation shown in Figure 4.

The transfer of biomass between megafauna and humans could have been achieved only by eating megafauna, but human body mass increases only until they reach maturity. Thereafter it remains approximately constant. There would have to be an enormous increase in the size of human population to account for the biomass transfer from these large animals, the transfer so massive that it caused their extinction. There is no indication of such a large surge in the growth of the population of *Phasmapithecus* interpreted as human population in Barnosky's publication (see Figures 1 and 2 based on his Fig. 2) and there is no indication of such a sudden surge in the real population data (Nielsen, 2017b). The growth of population before and after the extinction pulse was increasing *monotonically* by following hyperbolic distribution.

Figure 4, which is based on the original Fig. 3 in Barnosky's paper (Barnosky, 2008, p. 11544) suggests a nearly 100% transfer of the mass of megafauna to humans because the distribution representing the biomass of humans crosses the distribution representing the biomass of megafauna and even continues to increase above the biomass of megafauna. His comment, quoted earlier, also suggests a nearly 100% transfer of mass. This massive transfer complicates the balance sheet because it is hard to imagine that it could have been even achieved by a surge in the growth of population. Again, children and young adults would have to eat nothing else but megafauna. Human died certainly did not consist of megafauna, and children of these early humans did not eat exclusively megafauna. Certain body parts of megafauna, such as skin and bones, were not consumed as food, so again we have a problem with the balance sheet. We would also have to account for the loss of megafauna by natural causes, unrelated to human consumption.

These examples of fabricated "data" and of the claim that the extinction of megafauna was caused by humans, show how science can be degraded to fiction and how the misinformation is propagated even in peer-reviewed journals.

## Summary and conclusions

I have shown that the claimed correlation (Barnosky, 2008) between the growth of human population and the extinction of megafauna was based on invented "data," the "data" which do not describe the growth of human population but the growth of a phantom genus *Phasmapithecus*, the set of numbers created by Hern (1999) who never expected that they would be ever used as human population "data" (Hern, 2013). Furthermore, I have also shown that even these invented "data" do not support the hypothesis of the human-assisted extinction of megafauna because contrary to the published claim the essential correlation does not exist.

These two examples, one describing the fabricated data and the other describing the unsubstantiated claims about the extinction of megafauna, are disturbing. How could the research get so wrong?

How was it possible to publish fiction in the peer-reviewed scientific journals? How was it possible to publish invented data, while it was so obvious that they were invented and that they had nothing to do with the growth of human population? Even the Author of these fabricated data indicated in plain English that he was inventing them. How was it possible for the referee not to see it?

How was it possible for Barnosky to use these so obviously incorrect set of "data"? How was it possible to publish such clearly unsubstantiated evidence for the human-induced extinction of megafauna? How was it possible to accept for publication the claimed correlation, while it obviously did not exist? How was it possible to repeat this incorrect claim in another peer-reviewed publication supporting Barnowsky's unsubstantiated claim? Is it all just a sporadic aberration or is it a more serious systemic problem with the way research is conducted and with the way its results are accepted for publication?

Science is a self-correcting discipline but the process of self-correction starts, or at least it should start, with a rigorous and ruthless self-correction by individual scientists before they submit their research results for publication. Peer-reviewing by unbiased, well-informed and open-minded referees is added to check the scientific validity of the results submitted for publication. If this system is at any of these two stages upset, if individual scientists are tempted to promote their cherished and preconceived ideas, if referees are biased, ill-informed, prejudiced or simply lazy, any type of unsubstantiated claims can be published while progressive contributions will be rejected. Science will eventually recover; it will eventually correct the unsubstantiated claims but it might take a long time for this to happen.

Extensive correspondence with Tony Barnosky and Warren Hern is gratefully acknowledged.